\documentclass[rnote]{aa}  
\usepackage{graphicx,natbib,ulem}
\usepackage{txfonts}
\usepackage{color}

\begin{document}
  \titlerunning{Apsidal advance in SS 433?}
\authorrunning{ Bowler}
   \title{Apsidal advance in SS 433?}

   \subtitle{}

   \author{
          M.\ G.\ Bowler \
          }

  \offprints{M.G.Bowler \\  \email{m.bowler1@physics.ox.ac.uk}}
   \institute{University of Oxford, Department of Physics, Keble Road,
              Oxford, OX1 3RH, UK}
   \date{Received 15 October 2009; accepted 15 November 2009}

 
  \abstract
   {The Galactic microquasar SS\,433 launches oppositely-directed jets at speeds approximately a quarter of the speed of light. Both the speed and direction of the jets exhibit small fluctuations. A component of the speed variation has 13 day periodicity and the orbital phase at which its maximum speed occurs has advanced approximately 90$^{\circ}$ in 25 years. }
  {To examine the possibility that these variations are associated with a mildly eccentric orbit and conditions necessary to achieve this apsidal advance.}
   {The advance of the orbital phase for maximum speed is taken to be advance of the apses of the putative elliptical orbit. It is compared with calculations of the effects of tides induced in the companion and also with gravitational perturbations from the circumbinary disc. These calculations are made in the light of recent results on the SS 433 system.} 
  {The 13-day periodicity in the speed of the jets of SS 433 might be attributed to a mildly elliptical orbit, through periodic approaches of the donor and the compact object. Advance of the apses of such an elliptical orbit due to tidal effects induced in a normal companion looks to be too small; if caused by the circumbinary disc the mass of the inner regions of that disc is $\sim 0.15$$ M_\odot$.}
{\ldots}

   \keywords{Stars: individual: SS\,433 - stars: binaries: close}

   \maketitle
%

\section{Introduction}
The relativistic jets of SS 433 have a mean speed of about 0.26 $c$ but that speed is not constant;  it fluctuates with rms deviation $\sim 0.014$\,$c$. These fluctuations are rather symmetric between the jets and a component of the fluctuations has a period of about 13 days, the period of the binary orbit (Blundell and Bowler 2004, 2005; Blundell, Bowler \& Schmidtobreick 2007). The orbital phase of maximum speed of this component has advanced by $\sim 90$$^{\circ}$ in 25 years (Blundell, Bowler \& Schmidtobreick 2007). The mechanisms whereby the jet speed is modulated have not as yet been elucidated, but the periodic component suggests that variation of separation between the compact object and the companion is implicated; an elliptical orbit. Such an elliptical orbit has also been suggested as a reason for flares being associated with particular orbital phases claimed by Fabrika \& Irsmambetova (2002).

If the orbit of the binary is elliptical then the volume of the Roche lobe of the companion will reach a minimum once each orbit for eccentricity $e$ as small as 0.01  ( Bland \& Grindlay 1984). Periodic  approaches might increase the rate of transfer of material to the accretion disc or otherwise disturb its outer regions. It may be relevant that a period of higher than usual jet speeds set in on or just before JD 2453294 (Blundell, Bowler \& Schmidtobreick 2007), coinciding with an optical outburst and a radio flare (Schmidtobreick \& Blundell 2006). The ratio of the putative period of apsidal advance to the period of the system, 2796, is not atypical of massive close binaries where apsidal advance is induced by tidal effects (J I Katz private communication); for Y Cygni the ratio is 5791. The mass ratio of the SS 433 system $q= M_{\rm x}/M_{\rm c}$ and the masses $M_{\rm x}$ of the compact object and $M_{\rm c}$ of the companion have been much better established as a result of detailed studies of the H$\alpha$ structure which revealed the circumbinary disc (Blundell, Bowler \& Schmidtobreick 2008); the system is now known to be massive (roughly 40$ M_ \odot$). In the light of these new data, it seems worth while to examine conditions under which an elliptical orbit would precess at a rate of $\sim$ 90$^{\circ}$ in 25 years. Tidal distortion of the companion star by the compact object is one possible mechanism, another is the effect of gravitational perturbations due to the circumbinary disc itself. These are addressed in turn.

\section{Apsidal precession and tidal distortion}

The classic work on apsidal motion in binary systems as a result of tidal distortions is Sterne (1939).        In that treatment distortion is along the line joining the centres (equilibrium tides) and account is taken of the variation of tidal deformations with time. In this note his equation (14a) is applied, making the following assumptions. First, that the compact object (plus accretion disc) can be treated as a point mass (but in the context of SS 433 see Collins \& Newsom 1988). Secondly, that the companion co-rotates with the binary system and finally that terms in his equation (14) which involve the eccentricity $e$ (which enters as the square) can be ignored. Eq.(14a) of (Sterne 1939) then reads

\begin{equation}
\frac{\Omega_{\rm apses}}{\Omega_{\rm orbit}} =  k_{\rm 2} \left(\frac{a}{A}\right)^5 (16 q + 1)                                 
\end{equation}

Here $\Omega_{\rm apses}$ is the frequency of the induced apsidal advance and $\Omega_{\rm orbit}$ the orbital frequency. The structure constant $k_{\rm 2}$ is for the companion star (it would be equal to 0.0144 for a polytrope of $n=3$). The orbital speed of the circumbinary disc has revealed that $q$ is greater than 0.7 (Blundell, Bowler \& Schmidtobreick 2008); it probably lies in the range 0.7 - 1.0. The quantity  $a$  is the radius of the companion and $A$ is the semi-major axis of the binary orbit, equal to the separation of the two components for negligible ellipticity. For an orbit of period 13.08 days and periapsis advance of 90$^{\circ}$ in 25 years, the left hand side of Eq.(1) is  $3.58 \times 10^{-4}$. 

The ratio  $a/A$ is given by the length of time of total eclipse by the companion of an ideal luminous point in the orbit of the compact object. The length of time  $P_{\rm o}$ for which such an object is eclipsed when viewed in the orbital plane is

\begin{equation}
P_{\rm o} = \frac{13.08}{\pi} \frac{a}{A}
\end{equation}

where $P_{\rm o}$ is measured in days. The orbital plane is in fact inclined at 12$^{\circ}$ degrees to the line of sight and the actual eclipse does not last so long, rather

\begin{equation}
P  =  \frac{13.08}{\pi} \sqrt{ \left(\frac{a}{A}\right)^2 - 0.043}                                         
\end{equation}

The primary eclipses in SS 433 have been studied in both the optical and X-ray regions of the spectrum (see for example Goranskii, Esipov \& Cherepashchuk 1998, Cherepashchuk et al 2005) but in neither case is a pointlike object being eclipsed. The ideal eclipse time $P$ must however be close to a single day; for $P$ = 1 day, $a/A$ = 0.32. Another indication of the value of $a/A$ is obtained from the value of the mass ratio $q$ and the assumption that the companion fills its Roche lobe. The Roche lobe radius of the companion, in units of $A$, is a function only of $q$ (Eggleton 1983) and hence $\sim 0.4$. Thus Eq.(1) is easily satisfied for log $k_{\rm 2}$ $\sim -2$, provided that this value is compatible with the companion mass of 20 - 30 $M_{\odot}$ and the radius $\sim 2$ $10^{7}$ km. The latter corresponds to a surface gravity given by log $g$ $\sim$ 3 ($g$ in units of cm s$^{-2}$). Here there seems to be a problem - a normal star is too small if log $k_{\rm 2}$ $\sim -2$ or too concentrated after expanding to the required radius. The close OB stars in Y Cygni (Hill \& Holmgren 1995) provide an example where the masses and radii are known and log $k_{\rm 2}$ are -1.94, inferred from the apsidal precession. The stars each have mass $\sim 17$ $M_{\odot}$, radii $\sim 6$ $R_{\odot}$ and hence log $g$ $\sim 4$ - they are too small to match the properties of the companion in SS 433. The Y Cygni system is fairly young, but log $k_{\rm 2}$ becomes increasingly negative as the star slowly expands. Stellar models have been evolved by Claret \& Gimenez (1992) and later by Claret (2004) and their tables list both log $k_{2}$ and log $g$. The calculated structure constants $k_{2}$ which determine the effects of equilibrium tides on the precession of the apses of a binary orbit survive very well comparison with close binary systems (Claret \& Gimenez 1993). These evolved models have, for suitable masses, log $k_{\rm 2}$  $\sim -3$ as their radii approach the companion radius (which is close to that of its Roche lobe). Such a structure is too concentrated to drive the suggested apsidal precession in SS 433. The tables of Claret (2004) also list structures with log $k_{\rm 2}$ $\sim -2$ computed for old bloated stars, but for isolated stars the surface gravity is then log $g < 0$, corresponding to a surface radius much larger than the separation of the two components of the SS 433 binary; diffuse outer regions would have been stripped off. Thus the companion must have a distinctly peculiar structure if it is responsible for the precession of the apses at a rate of one complete rotation in 100 years. It has certainly endured a peculiar history.

\section{Apsidal precession induced by the circumbinary disc}

It is very simple to estimate the rate at which the apses of an elliptical orbit advance as a result of a circumbinary disc. It is not necessary to consider tidal distortions because the presence of a ring of matter, external to the binary itself, augments the attractive inverse square radial force on each component with a linear term directed outward. Very little is known about the circumbinary disc; the calculations below are for the simplest case of a ring of circumbinary material coplanar with the binary orbit. If the ring of matter orbits the centre of the binary at a radius $R$ and has (linear) mass density $\rho$, then the radial force per unit mass $F$ on a component at radius $r$ is given by

\begin{equation}
F=  - \frac{GM}{4r^{2}}   + \frac{2{\pi\rho}RG}{R^2}\frac{r}{R}                                                         
\end{equation}

for $q$=1 and component masses $M$, ignoring terms in $r^{3}/R^{3}$. The ratio $r/R$ is about 0.2; for present purposes Eq.(4) is adequate.

In the presence of the linear term the frequency of small radial oscillations about a radius $r_c$ is smaller than the rotational frequency at that radius and hence the apses advance. The vibrational frequency $\omega_r$ is given by

\begin{equation}
\omega_r^{2}= \frac{GM}{4r_c^3}(1-16x)
\end{equation}    

and the circular frequency $\omega_c$ by

\begin{equation}
\omega_c^{2}=\frac{GM}{4r_c^3}(1-4x)
\end{equation}

where the small quantity $x$ is given by $\frac{2\pi\rho R}{M}$$\frac{r_c^3}{R^3}$.

Then

\begin{equation}
\frac{\omega_c}{\omega_r}= 1+6x
\end{equation}

and one complete rotation of the ellipse ($\sim$100 years) takes 1/6$x$ 13.08 day periods. Hence if $M$ is 20 $M_{\odot}$ the mass of the inner part of the circumbinary disc must be $\sim$0.15$M_{\odot}$ if it is responsible for an advance of the periapsis of $\sim 90$$^{\circ}$ in 25 years.

\section{Conclusions}

This work is speculative in the sense that it is by no means clear that the observed oscillation of jet speeds with the period of the orbit is associated with the approach of the two components in an elliptical  orbit. The advance of the phase corresponding to maximum speed might however be explained by apsidal advance of the orbit.

 Both tidal distortion of the companion and the presence of the circumbinary disc will advance the periapsis; if the effects of tidal distortion of the companion are responsible then log $g\sim$3 must be accompanied by  log $k_{\rm 2}$ $\sim-2$ . Such a structure would be remarkable.

The rate of advance of periapsis induced by the circumbinary disc would match that of the jet speed maximum for a rather modest mass in the disc - approximately 0.15 $M_{\odot}$, which corresponds to about 1000 years of mass transfer.

\begin{acknowledgements}
I thank J I Katz for comments.
\end{acknowledgements}

\end{document}